# Signal and Image Processing with Sinlets

Alexander Y. Davydov

AlgoTerra LLC, 249 Rollins Avenue, Suite 202, Rockville, MD 20852, U.S.A.

E-mail: alex.davydov@algoterra.com

**Abstract**

This paper presents a new family of localized orthonormal bases – *sinlets* – which are well suited for both signal and image processing and analysis. One-dimensional sinlets are related to specific solutions of the time-dependent harmonic oscillator equation. By construction, each sinlet is infinitely differentiable and has a well-defined and smooth instantaneous frequency known in analytical form. For square-integrable transient signals with infinite support, one-dimensional sinlet basis provides an advantageous alternative to the Fourier transform by rendering accurate signal representation via a countable set of real-valued coefficients. The properties of sinlets make them suitable for analyzing many real-world signals whose frequency content changes with time including radar and sonar waveforms, music, speech, biological echolocation sounds, biomedical signals, seismic acoustic waves, and signals employed in wireless communication systems. One-dimensional sinlet bases can be used to construct two- and higher-dimensional bases with variety of potential applications including image analysis and representation.

FT – Fourier Transform

SD – Schwarzian Derivative

TDHO – Time-dependent Harmonic Oscillator

TDHOE – Time-dependent Harmonic Oscillator Equation

WKB – Wentzel-Kramers-Brillouin

DCR – Data Compression Ratio

## 1 Introduction

Many real-world applications are increasingly becoming signal-processing intensive, which stimulates the search for new more efficient signal representations and mathematical techniques. In the past, significant advances in signal processing have been made by reconsidering constraints and limitations imposed by the state of technology as it progresses as well as by reformulating old problems in new conceptual terms and thus bringing new mathematics into play to tackle them. The goal of this paper is to add a new mathematical instrument into the toolbox of contemporary signal processing by introducing a fresh family of localized orthonormal bases which we will refer to as *sinlets* or *sinlet bases*. Sinlet basis is a natural alternative to the Fourier transform (FT) for square-integrable transient signals with infinite support. Sinlets are closely related to some specific solutions of the equation that governs behavior of a time-dependent harmonic oscillator (TDHO), with the convenient feature of being well-localized and having infinite support at the same time. Furthermore, by construction, each sinlet has a well-defined and smoothly-varied instantaneous frequency known in analytical form, and decomposition of a square-integrable real transient into sinlets requires knowledge of only a countable set of real-valued coefficients. Approximate representation of a suitable transient signal can be achieved with much fewer coefficients than it is possible with the FT provided the same level of accuracy. These and other properties of sinlets to be discussed in this paper make them suitable for analyzing many real-world signals whose frequency content changes with time. The examples include (but are not limited to) radar and sonar waveforms, music, speech, biological echolocation sounds, biomedical signals, seismic acoustic waves, and waveforms utilized in wireless communication systems. Additionally, one-dimensional sinlet bases can be easily extended to turn into two- and higher-dimensional bases, with a wide variety of potential applications including image analysis and representation.

Over the past several decades, a number of new signal representations have been found, the best-known being wavelets [1], wavelet packets, cosine packets [2], Wilson bases [3], ridglets [4], curvelets [5], and shapelets [6]. However, none of these or other acknowledged signal representations can be considered as a 'silver bullet' well-suited for all cases of interest. Sinlets may not turn out to be an exception in this list. The detailed comparison analysis of sinlet bases versus other known transforms and representations is beyond the scope of this paper. Nevertheless, the convenience and elegance with which some important problems in signal processing can be reformulated and then solved using sinlets, as shown below, suggests that this novel signal representation is promising and deserves serious consideration.

The paper is organized as follows. Section 2 presents sinlets and their basic properties. Examples of sinlet-based techniques for signal denoising, non-uniform sampling, envelope construction, numerical differentiation, scaling, and efficient storage are given in Section 3. Finally, Section 4 provides examples of use of 2D sinlets in the field of image representation and analysis.

## 2 Sinlets – Localized Orthonormal Bases

### 2.1 Time-dependent Harmonic Oscillator

We begin with considering a classical harmonic oscillator with time-dependent frequency whose behavior is governed by the TDHO equation (TDHOE):

$$\ddot{x}(t) + \Omega^2(t) \cdot x(t) = 0 \tag{1}$$

Here overdot represents differentiation with respect to time and function $\Omega(t)$ can attain not only real but pure imaginary values as well so that $\Omega^2$ may become negative during some time interval. There is nothing contradictory or unphysical in allowing $\Omega^2(t)$ to become negative; this means only that, during the corresponding time interval, the point $x=0$ has the highest potential energy in contrast to being the point where the potential acquires its minimum - a typical situation. There exists no known procedure of obtaining an *exact* solution of the equation (1) for an arbitrarily *given* function $\Omega^2(t)$, although several standard techniques, such as the celebrated WKB method [7], yield good approximate solutions under certain favorable conditions. However, one can generate an unlimited number of pairs (TDHOE, its exact solution) using the technique described in detail in [8]. To obtain such a pair, we first write down the solution $x(t)$ in a specific (polar) form

$$x(t) = \frac{A}{\sqrt{\dot{\theta}(t)}} \cdot \cos[\theta(t)], \qquad (2)$$

where $A$ is a positive real constant, and $\theta(t)$ denotes a phase function which is assumed to have positive derivative everywhere ($\dot{\theta}(t) > 0, t \in \Re$). It can be easily verified that (2) is the solution of TDHOE (1) if and only if the following relation holds

$$\Omega^2(t) = \dot{\theta}^2 + \tfrac{1}{2} S(\theta), \qquad (3)$$

where $S(\theta)$ denotes the *Schwarzian derivative* (SD) given by [9]

$$S(\theta) = \frac{\dddot{\theta}}{\dot{\theta}} - \frac{3}{2}\left(\frac{\ddot{\theta}}{\dot{\theta}}\right)^2 \qquad (4)$$

We can then select a phase function $\theta(t)$ in analytical form such that it has continuous derivatives up to a third order at least along with the positive first-order derivative everywhere. Substituting this phase function into (2) and (3), we obtain a sought-for pair (TDHOE, its exact solution in analytical form).

Few remarks about SD are in order. The Schwarzian derivative appears in many apparently unrelated areas of mathematics, from differential equations to one-dimensional dynamics to differential geometry of curves (where it has a geometric interpretation in terms of curvature) to the theory of conformal mapping, etc. [9, 10]. The noteworthy feature of SD is that it is zero only for functions that are linear fractional transformations of the argument, i.e., $S(f) = 0$ if and only if $f(t) = \widehat{T}_{abcd}(t) = \frac{a \cdot t + b}{c \cdot t + d}$, where *a*, *b*, *c*, and *d* are constants with the determinant $ad - bc \neq 0$. Also recall that SD is invariant under linear fractional transformation of the function it operates upon: $S(\widehat{T}_{abcd}(f)) = S(f)$.

### 2.2 Definition of Sinlets

Now we are ready to employ the method described above for construction of a localized orthonormal basis from eigenfunctions of a specially designed TDHOE.

***Theorem.*** Let $\theta(t-t_0;\sigma)$ be a continuous function of variable $t \in \Re$ with parameters $t_0$ (center) and $\sigma$ (characteristic width) possessing the following properties:

1. The derivative $\dot\theta > 0$ and is limited for all $t \in \Re$;

2. $\theta(t-t_0;\sigma) = \begin{cases} 0, & t \to -\infty \\ 1, & t \to +\infty \end{cases}$ \hfill (5)

3. The SD exists and $S(\theta) \leq 0$ for all $t \in \Re$.

We call $\theta(t-t_0;\sigma)$ a *Mother-Phase* and form a family of her children as follows

$$\theta_n(t-t_0;\sigma) = \frac{\pi}{2} + \pi(n+1)\cdot\theta(t-t_0;\sigma), \qquad n = 0,1,2,\ldots \quad (6)$$

Then an orthonormal basis is given by a set of functions generated according to the formula

$$Sl_n(t-t_0;\sigma) = \sqrt{2\dot\theta(t-t_0;\sigma)} \cdot \sin[\pi(n+1)\theta(t-t_0;\sigma)], \quad n = 0,1,2,\ldots \quad (7)$$

These functions are centered around $t_0$ and are localized within a window of size determined by parameter $\sigma$.

***Proof.*** As shown previously, functions $x_n(t) = \dfrac{A}{\sqrt{\dot\theta_n}}\cos\theta_n$ with some arbitrary positive constant $A$ are the exact solutions of the corresponding TDHO equations

$$\ddot x_n + [\dot\theta_n^2 + \tfrac{1}{2}S(\theta_n)]\cdot x_n = 0, \quad n = 0,1,2,\ldots \quad (8)$$

In order for $x_n(t \to \pm\infty) \to 0$, the condition #2 of the Theorem is a necessity because it makes the cosine term in the formula for $x_n(t)$ vanish when $t \to \pm\infty$ while the derivative $\dot\theta_n$ (in the denominator) also tends to zero in the same limits. We have $\dot\theta_n^2 = \pi^2(n+1)^2\dot\theta^2$ and $S(\theta_n) = S(\theta)$, where, in the latest expression, we made use of the fact that SD remains unchanged under linear fractional transformation of a function it operates upon. Hence any of the equations (8) can be rewritten in the form

$$\ddot x_n - q(t)x_n + \lambda_n\chi(t)x_n = 0, \quad (9)$$

where

$$q(t) \equiv -\tfrac{1}{2} S(\theta) \geq 0$$
$$\chi(t) \equiv \dot{\theta}^2 > 0 \qquad (10)$$
$$\lambda_n \equiv \pi^2 (n+1)^2 > 0$$

Equation (9) with conditions (10) is recognized as the Sturm-Liouville system [11] with well-known property that its eigenfunctions $x_n(t)$ corresponding to distinct eigenvalues $\lambda_n$ are *orthogonal* relative to the *weight function* $\chi(t) > 0$, that is

$$\int_{-\infty}^{+\infty} x_n(t) \cdot x_m(t) \cdot \chi(t) dt = 0 \quad \text{if} \quad n \neq m \qquad (11)$$

Hence pairs of functions $\sqrt{\chi(t)} \cdot x_n(t) \propto \sqrt{\dot{\theta}(t)} \cdot \cos[\theta_n(t)]$ with distinct indexes are orthogonal relative to the weight *1*. To build an *orthonormal* set from these functions, we need to find normalization factors $A_n$:

$$\upsilon_n = A_n \cdot \sqrt{\dot{\theta}(t)} \cdot \cos[\theta_n(t)] = \frac{A_n}{\sqrt{\pi(n+1)}} \sqrt{\dot{\theta}_n(t)} \cdot \cos[\theta_n(t)] \qquad (12)$$

$$\int_{-\infty}^{+\infty} \upsilon_n^2 dt = \frac{A_n^2}{\pi(n+1)} \int_{-\infty}^{+\infty} \dot{\theta}_n \cdot \cos^2[\theta_n] dt = \frac{A_n^2}{\pi(n+1)} \int_{\pi/2}^{\pi(n+3/2)} \cos^2[\theta_n] d\theta_n =$$

$$\frac{A_n^2}{\pi(n+1)} \cdot \frac{\pi}{2} \cdot (n+1) = 1$$

Thus, we obtain $A_n = \sqrt{2}$ independently of *n* which yields an orthonormal set

$$\upsilon_n = \sqrt{2\dot{\theta}} \cdot \cos\left[\frac{\pi}{2} + \pi(n+1) \cdot \theta\right] = -\sqrt{2\dot{\theta}} \cdot \sin[\pi(n+1) \cdot \theta] \qquad (13)$$

The choice of signs of functions of an orthonormal set is not important because sign alterations do not change the orthonormality. Therefore, for simplicity sake, we omit the minus sign in r.h.s. of (13) and the statement of the Theorem immediately follows. ∎

We will refer to elements of the localized orthonormal set (7) as *sinlets* (for obvious reasons) and call the set itself a *sinlet basis*. The choice of Mother-Phase determines specific shapes of sinlets but does not affect their orthonormality as long as all the conditions of the Theorem are satisfied. Parameters $t_0$ and $\sigma$ determine the center of localization and the size of a window where sinlets are not negligibly small, respectively. Index $n$ equals the number of zero-crossings of the corresponding sinlet.

Note that formula (7) presents sinlets in a polar form making it easy to introduce *instantaneous amplitude* and *instantaneous frequency* of a sinlet. The amplitude turns out to be the same for all sinlets generated from the same Mother-Phase:

$$\rho_s = \sqrt{2 \cdot \dot{\theta}} \tag{14}$$

The sinlet's *angular instantaneous frequencies* are defined by

$$\omega_n(t) = \tfrac{d}{dt}[\theta_n(t)] = \pi(n+1)\dot{\theta}, \tag{15}$$

and sinlet's *instantaneous frequencies* are given by

$$\nu_n(t) = \tfrac{1}{2\pi}\omega_n = \tfrac{1}{2}(n+1)\dot{\theta} \tag{16}$$

Due to the requirement for Mother-Phase to have a positive derivative, the instantaneous frequencies are always positive quantities but become negligibly small outside the window of size controlled by parameter $\sigma$.

Two examples of sinlet bases are presented below for illustration purposes.

*Example 1.* Consider a particular form of the Mother-Phase:

$$\theta(t - t_0; \sigma) = \tfrac{1}{2}\left(1 + erf\left(\tfrac{t-t_0}{\sigma\sqrt{2}}\right)\right) \tag{17}$$

where $erf(x) = \tfrac{2}{\sqrt{\pi}} \int_0^x \exp(-z^2) dz$ denotes the *error function* [12]. The straightforward inspection verifies that all conditions of the Theorem are satisfied. The instantaneous frequencies have Gaussian shape

$$\nu_n = \frac{n+1}{2\sqrt{2\pi}\sigma} \exp\left(-\frac{(t-t_0)^2}{2\sigma^2}\right) \tag{18}$$

The explicit expression for the sinlets of this family is given by:

$$Sl_n(t) = \left(\tfrac{2}{\pi}\right)^{\tfrac{1}{4}} \sigma^{-\tfrac{1}{2}} \exp\left(-\tfrac{(t-t_0)^2}{4\sigma^2}\right) \sin\left(\tfrac{\pi}{2}(n+1)\left[1 + erf\left(\tfrac{t-t_0}{\sigma\sqrt{2}}\right)\right]\right) \tag{19}$$

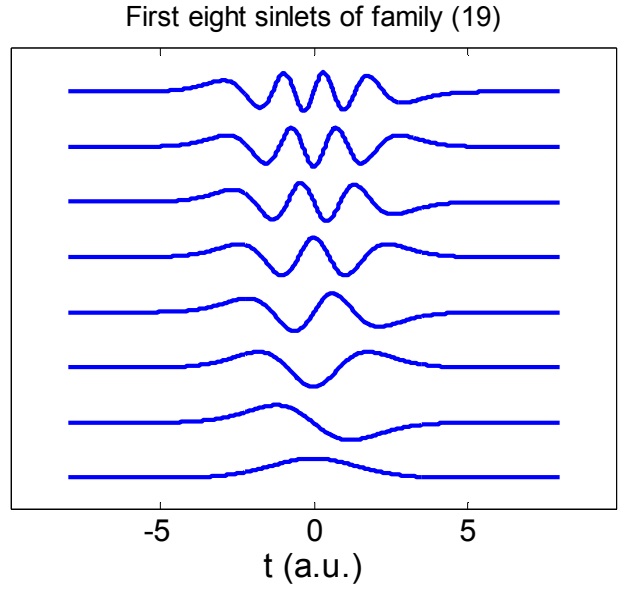

**Figure 1:** First eight sinlets of the family (19) with $t_0 = 0$ and $\sigma = 2$

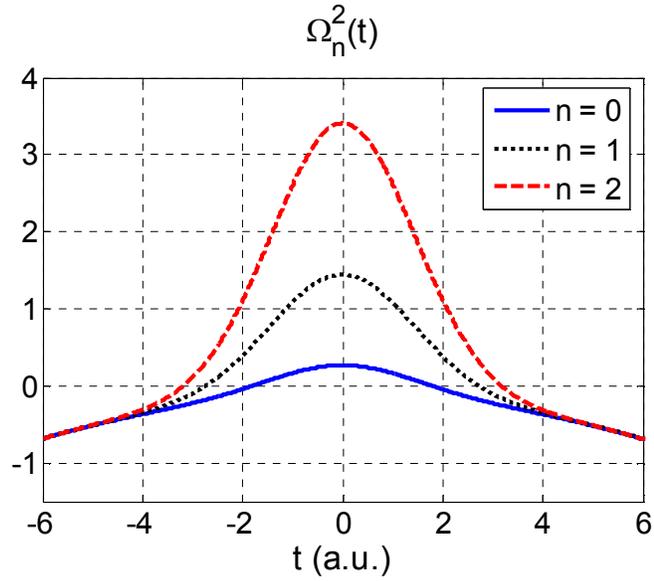

**Figure 2:** Functions $\Omega_n^2(t)$ for the first three sinlets of the family (19) with

$t_0 = 0$ and $\sigma = 2$

Figure 1 shows first eight sinlets of this family with some vertical offset for better representation. The corresponding functions $\Omega_n^2(t)$ are given by

$$\Omega_n^2(t) = \frac{1}{2\sigma^2}\left[\pi(n+1)^2 \exp(-\tfrac{(t-t_0)^2}{\sigma^2}) - \tfrac{(t-t_0)^2}{2\sigma^2} - 1\right], \qquad (20)$$

and first few of them are shown in Figure 2.

***Example 2.*** Consider a family of sinlets obtained from the Mother-Phase having the form of a logistic function

$$\theta(t - t_0;\sigma) = \left(1 + \exp\left(-\tfrac{t-t_0}{\sigma}\right)\right)^{-1} \qquad (21)$$

It is easy to check that all conditions of the Theorem are satisfied. The instantaneous frequencies have a soliton-like shape

$$v_n = \frac{n+1}{8\sigma \cosh^2[\tfrac{t-t_0}{2\sigma}]} \qquad (22)$$

The explicit expression for the sinlets of this family reads

$$Sl_n(t) = \sqrt{\tfrac{1}{2\sigma}}[\cosh(\tfrac{t-t_0}{2\sigma})]^{-1} \sin\{\pi(n+1)[1 + \exp(-\tfrac{t-t_0}{\sigma})]^{-1}\} \quad (23)$$

Figure 3 shows several first sinlets of this type. Figure 4 presents the corresponding instantaneous frequencies. The functions $\Omega_n^2(t)$ are given by

$$\Omega_n^2(t) = \tfrac{1}{4}\sigma^{-2}\left(\frac{\pi^2(n+1)^2}{4\cosh^4[\tfrac{t-t_0}{2\sigma}]} - 1\right) \qquad (24)$$

and first few of them are shown in Figure 5.

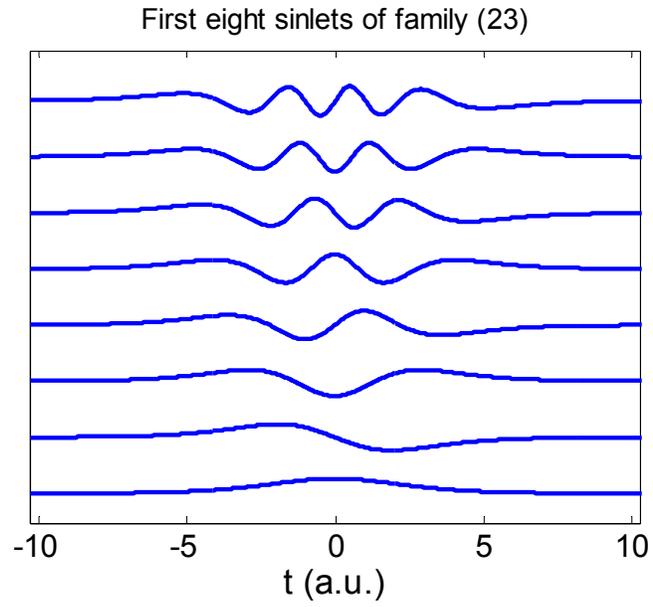

**Figure 3:** First eight sinlets of the family (23) with $t_0 = 0$ and $\sigma = 2$

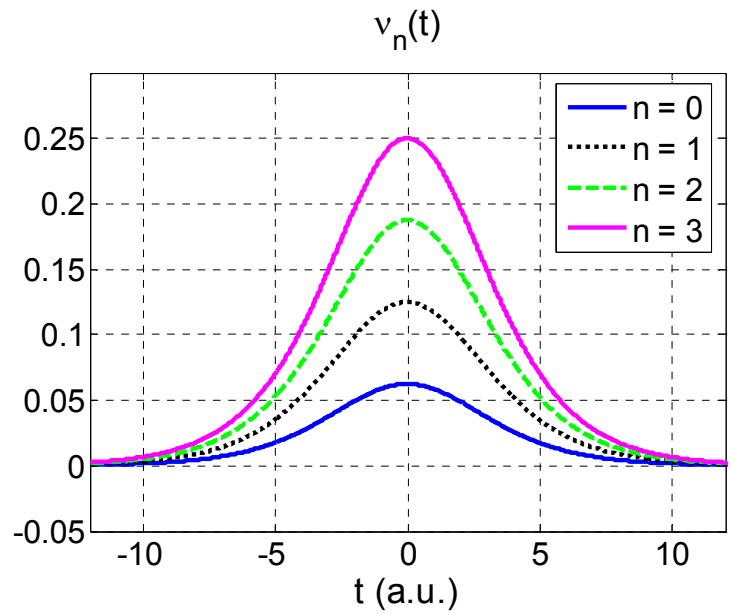

**Figure 4:** Instantaneous frequencies $\nu_n(t)$ for the first four sinlets of the family (23) with $t_0 = 0$ and $\sigma = 2$

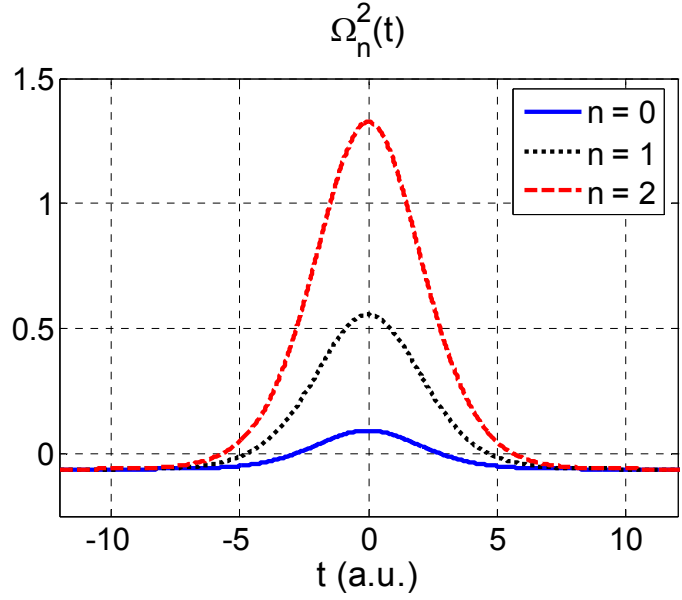

**Figure 5:** Functions $\Omega_n^2(t)$ for the first three sinlets of the family (23) with

$t_0 = 0$ and $\sigma = 2$

## 2.3 Basic Properties of Sinlets

### 2.3.1 Decomposition of Transient Bandpass Signals into Sinlets

Consider a square-integrable signal $u(t) \in L^2(\Re)$ occupying a frequency range from $\nu_{min} > 0$ to $\nu_{max}$ (with bandpass bandwidth $B = \nu_{max} - \nu_{min}$). To represent this signal by a weighted sum of sinlets, one has to optimally select (1) a center $t_0$, (2) an appropriate width parameter $\sigma$, and (3) a total number of sinlets $n_{max} + 1$ to be employed. These choices are signal-specific but some general rules of thumb can be employed as follows

(1) Choice of $t_0$:

$$t_0 \approx \frac{\int_{-\infty}^{+\infty} t \cdot u^2(t) dt}{\int_{-\infty}^{+\infty} u^2(t) dt}, \qquad (25)$$

provided that the integral in the nominator exists.

(2) Choice of $\sigma$:

$$\sigma \approx c \cdot \left[ \frac{\int\limits_{-\infty}^{+\infty} (t-t_0)^2 \cdot u^2(t)dt}{\int\limits_{-\infty}^{+\infty} u^2(t)dt} \right]^{\frac{1}{2}}, \qquad (26)$$

where constant $c \in [1, 2]$, and we have assumed that the integral in the nominator exists.

(3) Choice of $n_{max}$:

$$n_{max} = \left\lceil \frac{2\nu_{max}}{\dot{\theta}(t_{max} - t; \sigma)} \right\rceil - 1, \qquad (27)$$

where $t_{max}$ is the maximal value of the argument at which $u(t)$ is still non-negligible (in a sense specific for the application), i.e., $u(t) \approx 0$ for $t > t_{max}$, and $\lceil \cdot \rceil$ denotes the nearest integer towards infinity. For sinlet basis generated by the Mother-Phase (17), condition (27) yields

$$n_{max} = \left\lceil 2\sqrt{2\pi}\sigma\nu_{max} \exp\left(\frac{(t_{max}-t_0)^2}{2\sigma^2}\right) \right\rceil - 1 \qquad (28)$$

Similarly, for sinlets generated from the Mother-Phase (21), one obtains

$$n_{max} = \left\lceil 8\sigma\nu_{max} \cosh^2\left(\frac{t_{max}-t_0}{2\sigma}\right) \right\rceil - 1 \qquad (29)$$

The transient signal $u(t) \in L^2(\Re)$ can be now accurately represented as a sum

$$u(t) \cong \sum_{n=0}^{n_{max}} a_n \cdot Sl_n(t-t_0; \sigma), \qquad (31)$$

where the real-valued generalized Fourier coefficients $a_n$ are given by

$$a_n = \int\limits_{-\infty}^{+\infty} u(t) \cdot Sl_n(t-t_0; \sigma)dt, \quad n = 0, 1, \ldots, n_{max} \qquad (32)$$

Figure 6 shows an artificial bandpass signal and its single-sided Fourier amplitude spectrum and Figure 7 presents the decomposition of this signal into sinlets of family (23) along with comparison between original and reconstructed signals. Formulae (25), (26), and (29) were used with the following parameter values: $c = 1.5; \nu_{max} = 3.7 Hz; t_{max} = 3.8 s$.

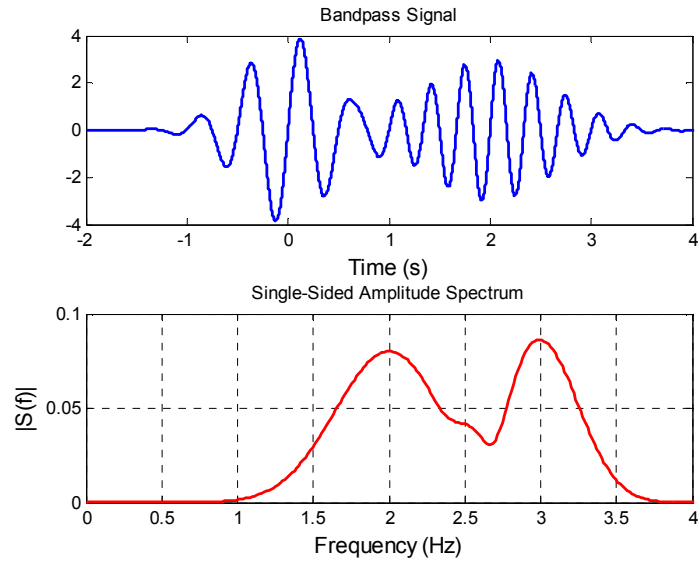

**Figure 6:** Artificial bandpass signal and its amplitude spectrum

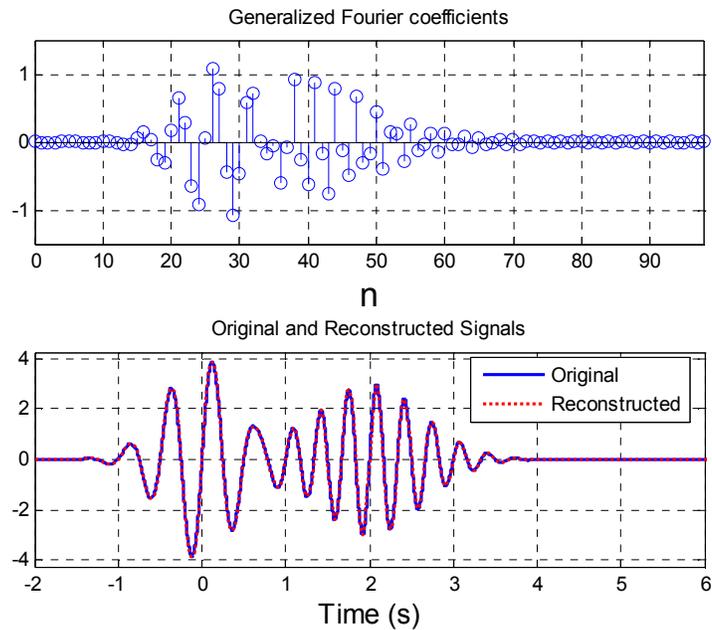

**Figure 7:** Decomposition of a bandpass signal into sinlets of family (23)

*2.3.2 Associated Orthonormal Bases: Coslets*

Let us define a localized orthonormal bases which are closely associated with sinlets and differ from them by replacement of sinusoids with cosinusoids:

$$Cl_n(t - t_0; \sigma) = \sqrt{2\dot{\theta}(t - t_0; \sigma)} \cdot \cos[\pi(n+1)\theta(t - t_0; \sigma)], \quad n = 0, 1, 2, ... \quad (33)$$

We will refer to functions of type (33) as *coslets*. The orthonormality of coslets can be checked by straightforward calculation. It is also easy to obtain the following relationship

$$D_{km} \equiv \int_{-\infty}^{+\infty} Sl_k(t) \cdot Cl_m(t) dt = \begin{cases} 0, & \text{if } k = m \\ \dfrac{1 - \cos(\pi(k + m + 2))}{\pi(k + m + 2)} + \dfrac{1 - \cos(\pi(k - m))}{\pi(k - m)}, & \text{if } k \neq m \end{cases} \quad (34)$$

Here we have assumed that both sinlets and coslets are generated from the same Mother-Phase and have the same values of parameters $t_0$ and $\sigma$, respectively. However, notice that the result (34) is independent from either $t_0$ or $\sigma$. Figure 8 shows first seven sinlets of family (23) with the corresponding coslets.

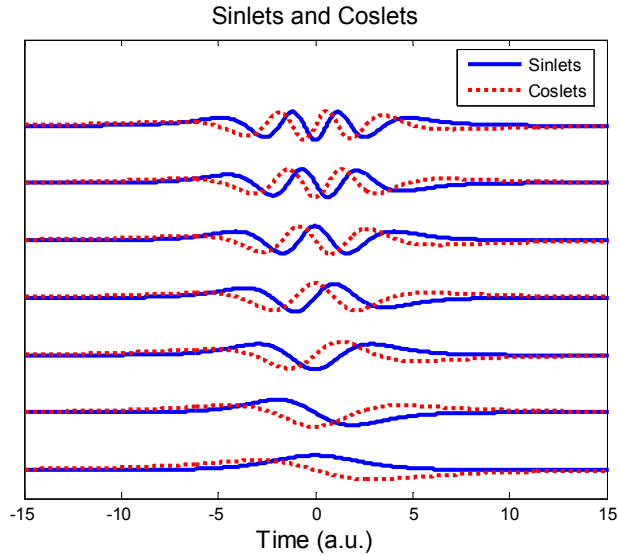

**Figure 8:** First seven sinlets and coslets of the family (23) with $t_0 = 0$ and $\sigma = 2$

It is convenient to introduce a joint complex-valued representation of sinlet and coslets as

$$\Psi_n(t-t_0;\sigma) \equiv Cl_n(t-t_0;\sigma) + i \cdot Sl_n(t-t_0;\sigma) = \sqrt{2\dot{\theta}} \cdot e^{i\pi(n+1)\theta} \quad (n = 0, 1, \ldots) \tag{35}$$

This representation is akin to analytic representation based on the Hilbert transform (HT) [14] and it can be employed for detection of signal's envelope as will be shown later. Figure 9 presents the first seven functions $\Psi_n$ based on sinlets of the family (23).

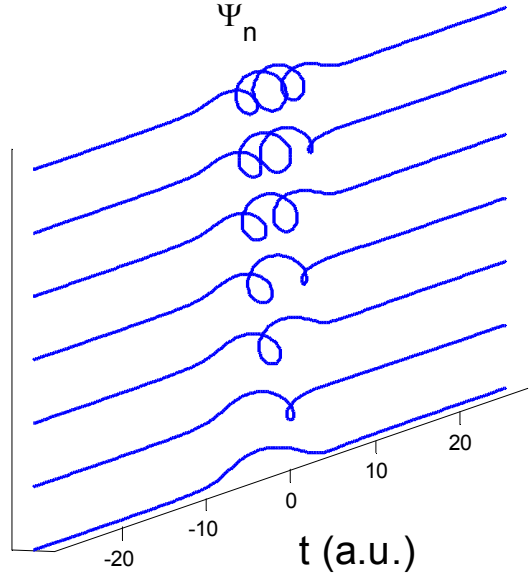

**Figure 9:** Joint complex-valued representation of sinlet and coslet bases

### 2.3.3 Connection between Decompositions into Sinlets and Coslets

Decomposing the same square-integrable signal $u(t)$ into sinlets and coslets, one can find the relationship between two sets of generalized Fourier coefficients. Indeed, assuming that signal $u(t)$ can be accurately represented by a finite number $N = n_{\max} + 1$ of sinlets and, alternatively, by a first $N$ coslets with the same parameters $t_0$ and $\sigma$, we obtain

$$\sum_{n=0}^{N-1} a_n \cdot Sl_n(t) \cong \sum_{m=0}^{N-1} b_m \cdot Cl_m(t) \tag{36}$$

After multiplying both sides of Eq. (36) by $Sl_k(t)$ and integrating, we arrive at

$$\vec{a} \cong D \cdot \vec{b}, \tag{37}$$

where $\vec{a}^T = (a_0, a_1, \ldots, a_{N-1})$, $\vec{b}^T = (b_0, b_1, \ldots, b_{N-1})$, and $N \times N$ matrix $D$ is given by (34). Similarly, after multiplying both sides of (36) by $Cl_k(t)$ and integrating, we obtain

$$\vec{b} \cong D^T \cdot \vec{a} \qquad (38)$$

Notice that here we employ $D^T$ instead of $D^{-1}$ as one might expect from Eq. (37). For odd values of $N$ (or even values of $n_{max}$), $\det(D) = 0$ and thus $D^{-1}$ does not exist. For even $N$ (odd $n_{max}$), $D^T \approx D^{-1}$ but using $D^T$ in (38) gives more precise results, especially for large $N$ because $\det(D) \to 0$ as $N$ increases. Numerical simulations indicate that both Eqs. (37) and (38) describe the relationships between generalized Fourier coefficients with good accuracy for arbitrary $N$.

## 3 Analyses of One-dimensional Transient Signals

In this Section, we present several applications of sinlets and coslets for processing and analysis of one-dimensional transient signals.

### 3.1 Sinlet Based Denoising

Consider a transient signal contaminated by an additive white Gaussian noise as shown in Figure 10. If we wish to remove the unwanted noise from the signal, we can simply decompose it into an appropriate small number of first sinlets (or coslets).

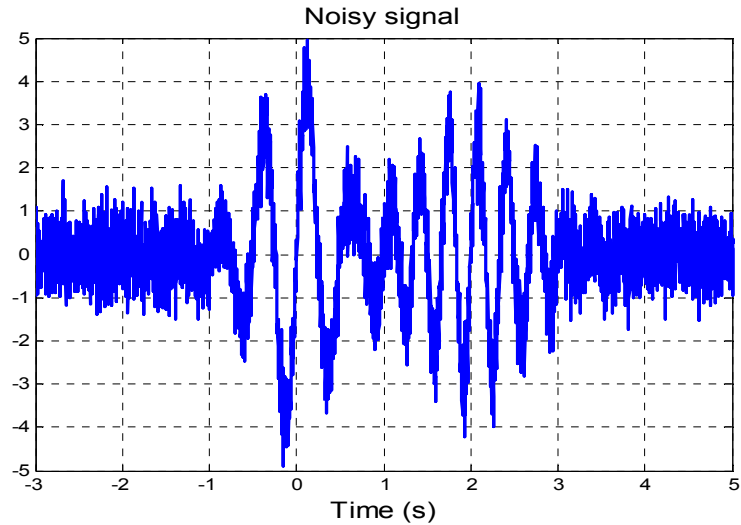

**Figure 10:** Artificial signal contaminated by an additive white Gaussian noise

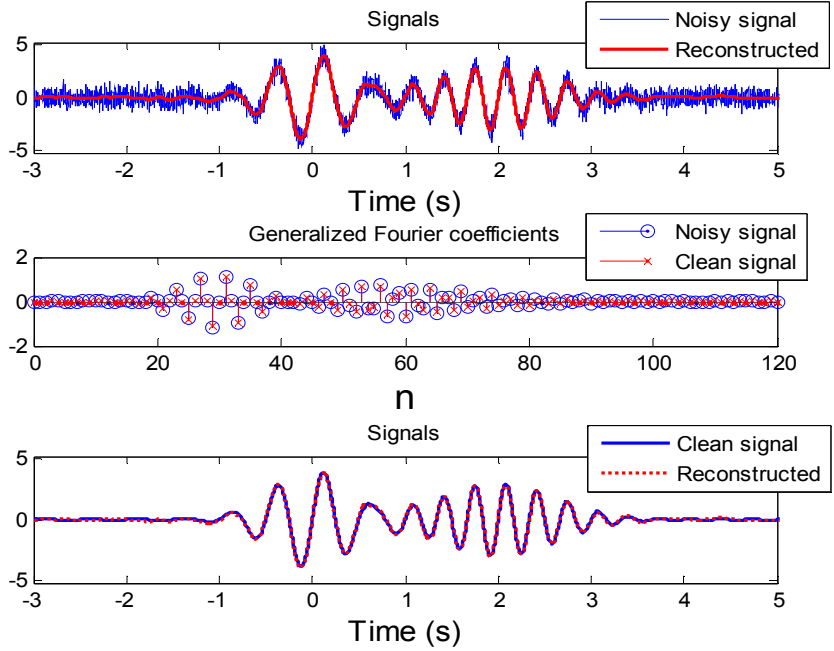

**Figure 11:** Results of the denoising by decomposition into sinlets of family (23)

Results of applying this technique to denoising of the artificial signal of Figure 10 are shown in Figure 11, see figure captions for details.

### 3.2 Non-uniform Sampling

Non-uniform (a.k.a. non-equidistant, irregular, staggered, uneven) sampling often is used intentionally or by necessity in various applications from radar and sonar to sensor networks to medical data collection to astronomical time series. Given a series of $N$ sample values $x_n$ sampled non-uniformly at times $t_n$ from a continuous-time transient signal $u(t)$ (i.e., $x_n \equiv u(t_n), n = 1, \ldots, N$), how can we find the accurate representation of this continuous signal, $u(t)$?

Assume that the signal under consideration can be approximated by a linear superposition of $K$ sinlets with parameters $t_0$ and $\sigma$:

$$u(t) \cong \sum_{k=0}^{K-1} a_k \cdot Sl_k(t - t_0; \sigma) \qquad (39)$$

Now the problem boils down to finding $\vec{a}^T = (a_0, a_1, \ldots, a_{K-1})$ from the system of linear equations

$$\vec{x} = F \cdot \vec{a}, \qquad (40)$$

where $\vec{x}^T = (x_1, x_2, \ldots, x_N)$, and $N \times K$ matrix $F$ is given by

$$F = \begin{pmatrix} Sl_0(t_1 - t_0; \sigma) & Sl_1(t_1 - t_0; \sigma) & \cdots & Sl_{K-1}(t_1 - t_0; \sigma) \\ Sl_0(t_2 - t_0; \sigma) & Sl_1(t_2 - t_0; \sigma) & \cdots & Sl_{K-1}(t_2 - t_0; \sigma) \\ \vdots & \vdots & \cdots & \vdots \\ Sl_0(t_N - t_0; \sigma) & Sl_1(t_N - t_0; \sigma) & \cdots & Sl_{K-1}(t_N - t_0; \sigma) \end{pmatrix} \quad (41)$$

Assuming that $K < N$, matrix $F$ has a rank $K$ (it contains $K$ linearly independent columns) and hence the system (40) is overdetermined. The approximate solution, in the least squares sense, is given by the formula [13]

$$\vec{a} = (F^T \cdot F)^{-1} \cdot F^T \cdot \vec{x} \quad (42)$$

Figure 12 presents an example of reconstructing an artificial signal from non-uniform samples randomly chosen from interval [-5, 5] using continuous uniform distribution ($N = 150$, $K = 32$). No additive noise was assumed during 'sampling' procedure. The reconstructed signal practically coincides with the original one.

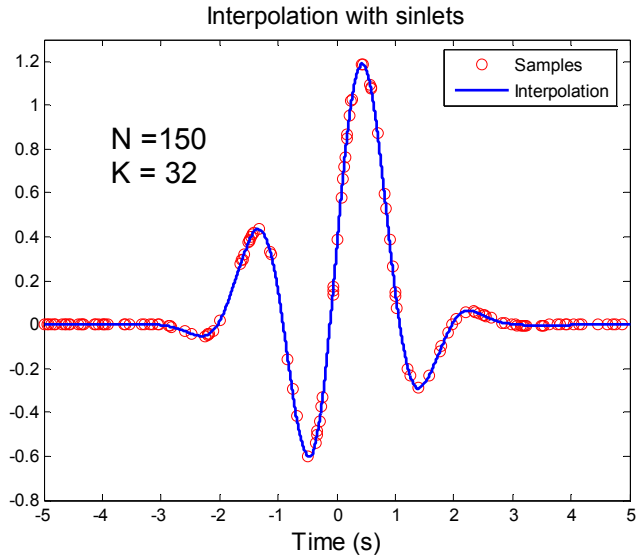

**Figure 12:** Reconstruction of signal from non-uniform samples using sinlets of family (23) ($N = 150$; $K = 32$)

Figure 13 shows the results of signal reconstruction for the case when samples are contaminated by additive white Gaussian noise ($N = 300$; $K = 32$). The reconstructed signal is in good agreement with the original one.

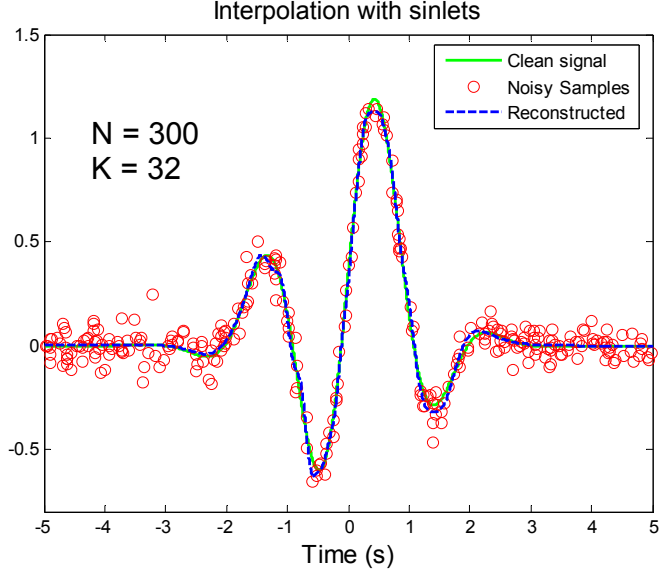

**Figure 13:** Reconstruction of signal from non-uniform noisy samples using sinlets of family (23) ($N = 300$; $K = 32$)

### 3.3 Envelope Detection

Envelope detection is one of the most common tasks in signal processing applications. HT is often used for this purpose. Here we present a novel method for calculating envelope of a square-integrable transient signal $u(t)$ based on decomposition into sinlets or coslets. Let $u(t) \cong \sum_{n=0}^{N-1} a_n \cdot Sl_n(t - t_0; \sigma)$. Then a real-valued envelope of the signal is given by

$$\text{Envelope } [u(t)] \cong \left| \sum_{n=0}^{N-1} a_n \cdot \Psi_n(t - t_0; \sigma) \right| = \left| \sqrt{2\dot{\theta}} \cdot \sum_{n=0}^{N-1} a_n e^{i\pi(n+1)\theta} \right| \quad (43a)$$

Alternatively, if $u(t) \cong \sum_{n=0}^{N-1} b_n \cdot Cl_n(t - t_0; \sigma)$, one can write

$$\text{Envelope } [u(t)] \cong \left| \sum_{n=0}^{N-1} b_n \cdot \Psi_n(t - t_0; \sigma) \right| = \left| \sqrt{2\dot{\theta}} \cdot \sum_{n=0}^{N-1} b_n e^{i\pi(n+1)\theta} \right| \quad (43b)$$

Formulae (43a) and (43b) give almost identical results. For signals without higher-frequency noise, these envelopes practically coincide with the HT-based envelope, as shown in Figure 14.

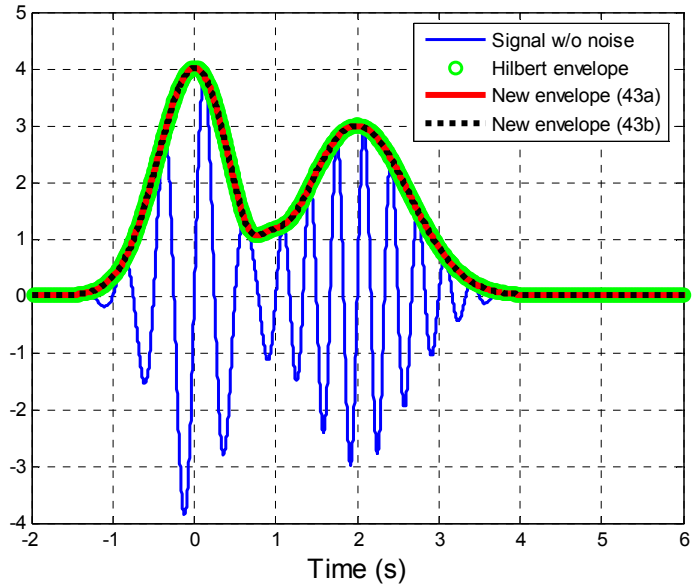

**Figure 14:** Comparison between three methods of envelope detection for noise-free artificial signal

However, if signals are contaminated by an additive white Gaussian noise, numerical simulations indicate that the novel method of calculating envelope yields more robust results than the HT-based approach, as Figure 15 clearly illustrates.

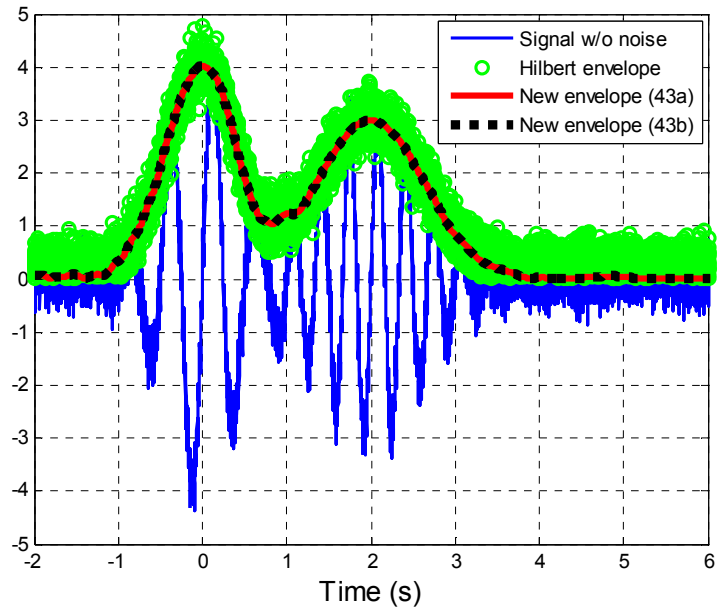

**Figure 15:** Comparison between three methods of envelope detection for noisy artificial signal

## 3.4 Numerical Differentiation

In many applications, it is desired to estimate the derivative of a function given its noisy values. This is an ill posed problem since small perturbations of the function result in large errors in its estimated derivative. Here we show how the new method based on decomposition of a transient square-integrable signal into sinlets can yield very accurate and robust estimation of signal's derivative even in presence of significant additive noise (which is not considered as part of the signal and hence must be filtered out).

We start by assuming that, in the absence of noise, the signal we want to differentiate can be accurately represented as a weighted sum of first $N$ sinlets $u(t) \cong \sum_{n=0}^{N-1} a_n \cdot Sl_n(t-t_0;\sigma)$, and then find the signal derivative

$$\dot{u}(t) \cong \sum_{n=0}^{N-1} a_n \left( \frac{\ddot{\theta}}{2\dot{\theta}} \cdot Sl_n(t-t_0;\sigma) + \pi(n+1)\dot{\theta} \cdot Cl_n(t-t_0;\sigma) \right) \qquad (44)$$

Hence, once the generalized Fourier coefficients $a_n$ are found, they can be used immediately to calculate the signal derivative. General formula (44) can be rewritten in especially simple form if we use sinlets (and associated coslets) of the family (19):

$$\dot{u}(t) \cong \sum_{n=0}^{N-1} a_n \left( \frac{t_0-t}{2\sigma^2} \cdot Sl_n(t-t_0;\sigma) + \sqrt{\frac{\pi}{2}}(n+1)\sigma^{-1} \exp\left(-\frac{(t-t_0)^2}{2\sigma^2}\right) \cdot Cl_n(t-t_0;\sigma) \right) \quad (45)$$

Figure 16 shows an artificial noisy signal and Figure 17 presents its smooth derivative estimated with the Eq. (45).

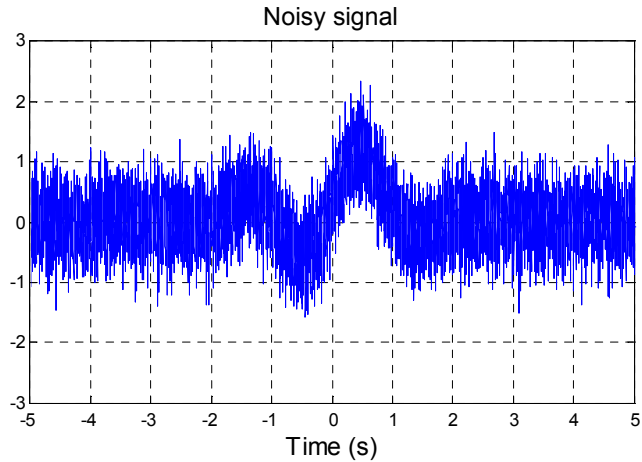

**Figure 16:** Artificial signal contaminated by additive noise. The derivative of the noise-free signal is wanted.

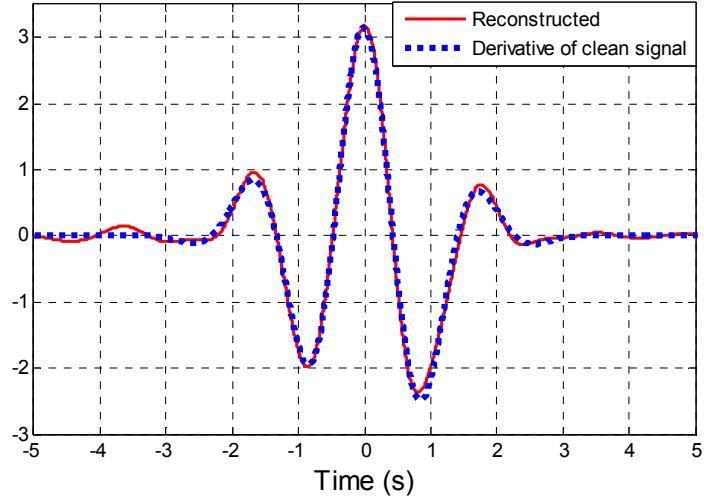

**Figure 17:** Extracting smooth derivative from noisy signal depicted in Fig. 16

Higher-order derivatives of transient signals can be estimated along similar lines.

### 3.5 Waveform Scaling and Efficient Storage

One of the useful features of sinlets is their scaling property expressed as

$$\sqrt{\alpha} \cdot Sl_n(\alpha(t-t_0); \sigma) = Sl_n(t-t_0; \alpha^{-1}\sigma), \quad n = 0, 1, \ldots, \tag{46}$$

where $\alpha$ is an arbitrary positive real number. This property comes in handy when one wants to emulate Doppler distortion of waveforms transmitted by radar or sonar. Indeed, for a point target moving with a constant velocity, the returned signal depends on two parameters: (i) a round-trip time delay and (ii) a Doppler scale caused by relative velocity between the radar/sonar and the target. Assuming both transmitted and echo signals are normalized to have the same energy, one can express the returned signal as [15]

$$w_r(t) = \sqrt{\alpha} \cdot w_{tr}[\alpha(t-\tau)], \tag{47}$$

where $w_{tr}(t)$ is the transmitted signal, $\alpha$ denotes the Doppler factor, and $\tau$ is the propagation time delay:

$$\alpha = \frac{c-\upsilon}{c+\upsilon}, \tag{48}$$

$$\tau = \frac{2D}{c} \tag{49}$$

Here $c$ is the velocity of the signal propagation, $v$ is the radial velocity between the radar/sonar and the target, and $D$ is the distance to the target. Assume that the first $N$ generalized Fourier coefficients $\{a_n, n = 0,1,\ldots,N-1\}$ provide an accurate representation of the transmitted waveform $w_{tr}(t)$ in sinlet basis $\{Sl_n(t-t_0;\sigma), n=0,1,\ldots\}$. Then, employing (46) and (47), one can express the Doppler-distorted and delayed received waveform as

$$w_r(t) \cong \sum_{n=0}^{N-1} a_n \cdot Sl_n(t - t_0 - \tau; \alpha^{-1}\sigma) \qquad (50)$$

Thus sinlet bases not only allow for compact representation of transient signals but provide means to easily emulate the distorting of these signals caused by the Doppler effect. This feature can be employed in data bases of target signatures for a multitude of civilian and military applications.

## 4 Analyses of Images with 2D Sinlets

### 4.1 Definition of *n*D Sinlet Bases

One-dimensional sinlet bases can be easily generalized to form two- and higher-dimensional localized orthonormal bases.

*Definition.* Let $\{\sigma_s > 0, s = 1,\ldots,n\}$ and $\{x_s^{(0)} > 0, s = 1,\ldots,n\}$ be two sets of positive real numbers and arbitrary real numbers, respectively. Choosing a particular one-dimensional sinlet basis, we can form an *n*-dimensional orthonormal basis from it as follows:

$$\Phi_{k_1 k_2 \ldots k_n} = Sl_{k_1}(x_1 - x_1^{(0)};\sigma_1) \cdot Sl_{k_2}(x_2 - x_2^{(0)};\sigma_2) \cdots Sl_{k_n}(x_n - x_n^{(0)};\sigma_n), \qquad (51)$$

where $k_i = 0,1,\ldots$ for any $i = 1,2,\ldots,n$.

The proof of the orthonormality of functions $\Phi_{k_1 k_2 \ldots k_n}$ is straightforward and is omitted for space sake. Figure 18 depicts several first two-dimensional orthonormal functions generated from the one-dimensional sinlets of family (23) with $x_1^{(0)} = x_2^{(0)} = 0$ and $\sigma_1 = \sigma_2 = 0.75$.

One can also construct *n*D localized orthonormal bases using one-dimensional coslets instead of sinlets in a similar fashion.

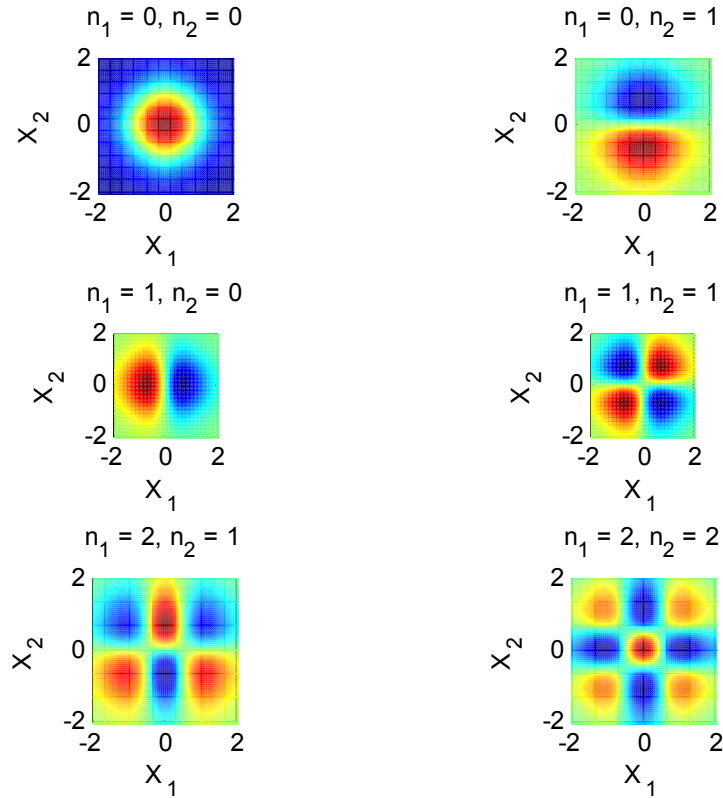

**Figure 18:** Examples of functions forming a two-dimensional orthonormal basis from one-dimensional sinlets of family (23) with $x_1^{(0)} = x_2^{(0)} = 0$ and

$$\sigma_1 = \sigma_2 = 0.75$$

### 4.2 Image Representation with 2D Sinlets

Representation of images is the first application that comes to mind when thinking about possible utilizations of 2D sinlet bases. Figures 19, 20, and 21 show examples of such image representations and clearly demonstrate the property of sinlets to facilitate image lossy compression. Notice that the least *data compression ratio* (DCR), defined as a ratio of compressed and uncompressed sizes, is achieved for the image with large uniform areas depicted in Figure 20. The detailed comparison analysis of sinlet-based and other known image compression techniques is beyond the scope of this paper and will be published elsewhere.

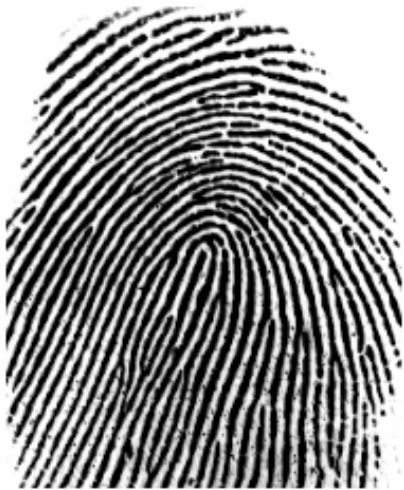 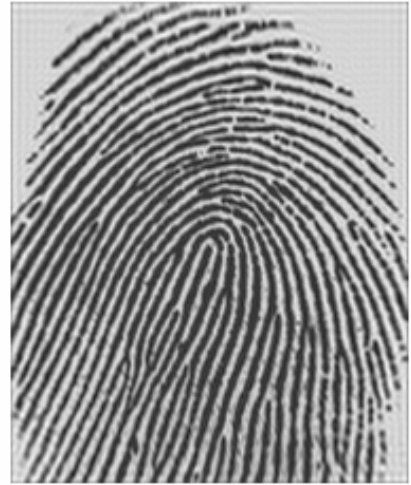

**Figure 19:** Fingerprint (DCR = 0.6162)

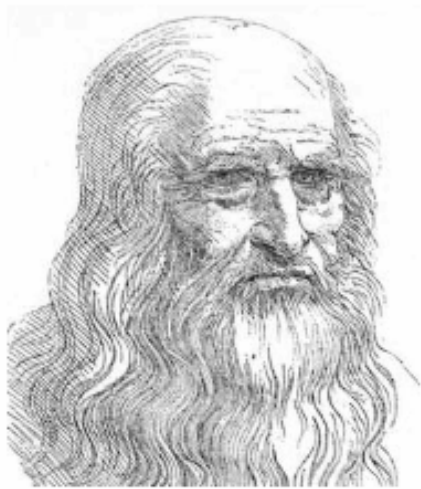 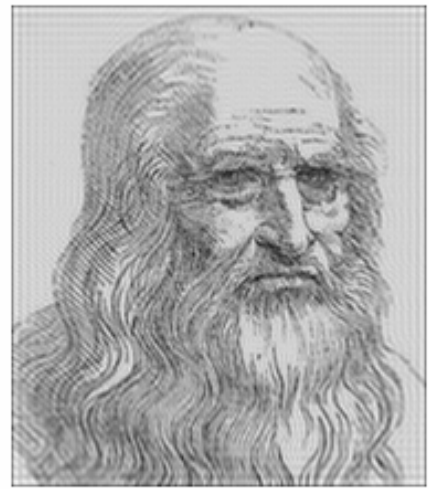

**Figure 20:** Self-portrait of Leonardo da Vinci (DCR = 0.4644)

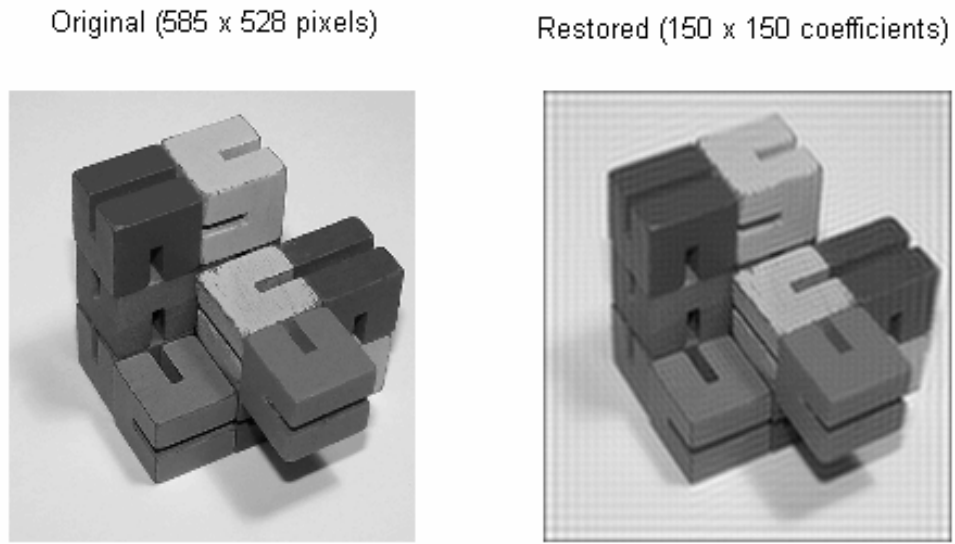

**Figure 21:** Three-dimensional object (DCR = 0.073)

## Acknowledgments

I am grateful to Prof. Ram Narayanan for insightful discussions. This work was supported by internal R&D funds from AlgoTerra LLC.